\documentclass[11pt]{article}
\usepackage{graphicx}
\usepackage{latexsym,graphicx}
\usepackage{float}
\usepackage{amsmath}
\usepackage{amsfonts}
\usepackage{amssymb}
\usepackage{epstopdf}
\usepackage{color}
\DeclareGraphicsExtensions{.eps}
\usepackage[left=2.5cm,top=2.5cm,right=2.5cm,bottom=2.5cm]{geometry}

\catcode `\@=11
\catcode `\@=12
\begin{document}
	\begin{center}
		\large{\bf{Transit cosmological models in Myrzakulov $F(R,T)$ gravity theory}} \\
		\vspace{5mm}
		\normalsize{ Dinesh Chandra Maurya$^{1}$,  Ratbay Myrzakulov$^{2}$}\\
        \vspace{5mm}
        \normalsize{$^{1}$ Centre for Cosmology, Astrophysics and Space Science, GLA University, Mathura-281 406,
	     Uttar Pradesh, India.}\\
         \vspace{5mm}
         \normalsize{$^{2}$ Eurasian International Centre for Theoretical Physics and Department of General \& Theoretical Physics, Eurasian National University, Astana 010008, Kazakhstan.}\\
         \vspace{2mm}
         {$^{1}$Email:dcmaurya563@gmail.com}\\
         \vspace{2mm}
         {$^{2}$Email:rmyrzakulov@gmail.com}\\
     \end{center}
         \vspace{5mm}
	%\date{}
	%\maketitle
	%%%%%%%%%%%%%%%%%%%%%%%%%%%%%%%%%%%%%%%%%%%%%%%%%%%%%%%%%%%%%%%%%%%%%%%%%%%%%%%%%%%%%%%%%%%%
	\begin{abstract}
		In the present paper, we investigate some exact cosmological models in Myrzakulov $F(R,T)$ gravity theory. We have considered the arbitrary function $F(R, T)=R+\lambda T$ where $\lambda$ is an arbitrary constant, $R, T$ are respectively, the Ricci-scalar curvature and the torsion. We have solved the field equations in a flat FLRW spacetime manifold for Hubble parameter and using the MCMC analysis, we have estimated the best fit values of model parameters with $1-\sigma, 2-\sigma, 3-\sigma$ regions, for two observational datasets like $H(z)$ and Pantheon SNe Ia datasets. Using these best fit values of model parameters, we have done the result analysis and discussion of the model. We have found a transit phase decelerating-accelerating universe model with transition redshifts $z_{t}=0.4438_{-0.790}^{+0.1008}, 0.3651_{-0.0904}^{+0.1644}$. The effective dark energy equation of state varies as $-1\le\omega_{de}\le-0.5176$ and the present age of the universe is found as $t_{0}=13.8486_{-0.0640}^{+0.1005}, 12.0135_{-0.2743}^{+0.6206}$ Gyrs, respectively for two datasets.
	\end{abstract}
	\smallskip
	\vspace{5mm}
	%\date{}
	%\maketitle
	\vspace{1cm}
	
	\tableofcontents
	%%%%%%%%%%%%%%%%%%%%%%%%%%%%%%%%%%%%%%%%
	\section{Introduction}
	%%%%%%%%%%%%%%%%%%%%%%%%%%%%%%%%%%%%%%%%
	
	The universe underwent two episodes of accelerated expansion at early and late times of the cosmological evolution, according to the conventional paradigm of cosmology, which is based on a growing amount of observable data. Although the cosmological constant might be the best explanation for the late-time acceleration, the possibility that the acceleration is dynamic in nature and the presence of some potential tensions may call for a revision of our understanding—something that is unquestionably necessary for early time acceleration. There are primarily two paths one could take in order to accomplish this. The first is to build extended gravitational theories that, although having general relativity as a specific limit, can generally offer more degrees of freedom to adequately describe the evolution of the universe \cite{ref1,ref2}. The second approach is to modify the conventional particle physics model and take general relativity into account. This involves assuming that the universe contains additional matter in the form of dark energy \cite{ref3,ref4} and/or inflation fields \cite{ref5}. Keep in mind that the first approach has the extra theoretical benefit of may be leading to an improved \cite{ref6,ref7}.\\
	
	One can begin building gravitational modifications from the Einstein-Hilbert action, that is, from the curvature description of gravity, and extend it appropriately, as in the cases of Lovelock gravity \cite{ref8,ref9}, $F(R)$ gravity \cite{ref10}, and $F(G)$ gravity \cite{ref11,ref12}. He may also examine torsional modified gravities, such as $F(T)$ gravity \cite{ref13,ref14,ref15}, $F(T, T_{G})$ gravity \cite{ref16}, etc., starting with the analogous, teleparallel formulation of gravity in terms of torsion \cite{ref17,ref18}. Cosmologists are interested in $f(T)$ teleparallel gravity, one of the intriguing modified theories of gravity. The study of modified Teleparallel $f(T)$ gravity, where $T$ is the torsion scalar, was driven by the generalization of $f(R)$ gravity, where $R$ is the Ricci scalar. To characterize the effects of gravitation in terms of torsion rather than curvature, \cite{ref19}-\cite{ref22} employed the curvatureless Weitzenb\"{o}ck connection in teleparallel gravity, as opposed to the traditional torsionless Levi-Civita connection in general relativity.\\
	
	The linear forms of $f(T)$ lead to a teleparallel gravity equivalent to general relativity (TEGR) \cite{ref23}. Nonetheless, there are disparities in the physical interpretations of the two theories of gravity, $f(T)$ and $f(R)$. In $f(T)$ gravity, the torsion scalar $T$ just comprises the first-order derivatives of the vierbeins, but in $f(R)$ gravity, the second-order derivatives of the metric tensor are contained in the Ricci scalar $R$. This means that, as opposed to other modified theories of gravity, the exact solutions of cosmological models in $f(T)$ gravity may be readily found. $f(T)$ gravity is a straightforward modified theory of gravity, however there aren't many precise solutions suggested in the literature. In isotropic and anisotropic spacetime, some cosmological models \cite{ref24, ref25, ref26} with power-law solutions have been discovered in the literature. Some cosmologists have studied some exact solutions of the cosmological models in \cite{ref19, ref27, ref28} for static spherically symmetric spacetime and Bianchi type-I spacetime. When compared to other modified theories of gravity, the analysis of cosmic situations in $f(T)$ gravity is straightforward. Consequently, a large number of cosmological scenarios, including the big bounce \cite{ref29, ref30, ref31, ref32}, inflationary model \cite{ref33}, and late time cosmic acceleration \cite{ref34, ref35, ref36}, are studied using $f(T)$ gravity theory. In the field of $f(T)$ gravity, there have been recent developments including spherical and cylindrical solutions (\cite{ref37}), conformally symmetric traversable wormholes (\cite{ref38}), and noether charge and black hole entropy (\cite{ref39}). Recently, we have discussed and reconstructed some $\Lambda$CDM cosmological models in $f(T)$ gravity \cite{ref40}-\cite{ref43}.\\	
	
	Additionally, nonmetricity could be used to create gravitational alterations \cite{ref44}. Furthermore, altering the fundamental geometry itself might give rise to an intriguing class of modified gravity; this could include, for example, Finsler or Finsler-like geometries \cite{ref45}-\cite{ref48}. The non-linear connection's potential to introduce additional degrees of freedom and make the gravitational modification phenomenologically interesting is one of the framework's intriguing features \cite{ref49,ref50}. This feature was also obtained through the use of a different theoretical framework for metric-affine theories \cite{ref51}-\cite{ref55}.\\
	
	In \cite{ref56}, R. Myrzakulov found an intriguing gravitational modification called the $F(R, T )$ gravity. Both curvature and torsion are dynamical fields associated with gravity in this theory because one makes use of a particular but non-special connection. Because of this, the theory has additional degrees of freedom originating from both the non-special connection and the arbitrary function in the Lagrangian. The theory belongs to the class of Riemann-Cartan theories, which are part of the broader category of metric theories with affine connections \cite{ref57,ref58}. A few of the theory's applications were examined in \cite{ref56} and \cite{ref59}–\cite{ref62}. Specifically, \cite{ref56} addressed certain theoretical concerns; \cite{ref59} examined energy conditions; \cite{ref60} examined theoretical relationships with various scenarios; \cite{ref61} examined Noether symmetries; and \cite{ref62} examined neutron star theory.	Recently, in \cite{ref63} have analyzed the resultant cosmology of such a framework and to compute the evolution of observable quantities like the effective dark energy equation-of-state parameter and density parameters. By expressing the theory as a deformation from both general relativity and its teleparallel counterpart, they have examined the cosmological behavior with an emphasis on the connection's effect by employing the mini-super-space approach. The observational constraints on Myrzakulov $F(R,T)$-gravity have been investigated in \cite{ref64}. Various Metric-Affine Myrzakulov Gravity Theories and its applications are discussed in \cite{ref65}-\cite{ref71}.\\
	
	Motivated by the above discussions, in this paper, we investigate a spatially flat, isotropic and homogeneous spacetime universe in Myrzakulov $F(R, T)$ Gravity. The paper is organized as follows. In Section 2, we give a brief review of the Myrzakulov $F(R,T)$ gravity theory. The cosmological solution for the particular linear case $F(R,T)=R+\lambda T$ are given in Section 3. Observational constraints for the model are studied in Section 4. The result analysis and discussions are presented in Section 5. The age of the universe is considered in Section 6.  The last Sec. 7 is devoted to conclusions.

	%=======================================================================================================================
	\section{Myrzakulov $F(R,T)$ gravity and field equations}
	%=======================================================================================================================
	
	To explore the cosmological properties of the universe in Myrzakulov $F(R,T)$ gravity, we consider the flat FRW space-time described by the metric	
	\begin{equation}\label{eq1}
		ds^{2}=-dt^{2}+a^{2}(t)(dx^{2}+dy^{2}+dz^{2}),
	\end{equation}
	where $a = a(t)$ is the scale factor. The orthonormal tetrad components $e_{i}(x^{\mu})$ are related to the	metric through
	\begin{equation}\label{eq2}
		g_{\mu\nu}=\eta_{ij}e_{\mu}^{i}e_{\nu}^{j},
	\end{equation}
	where the Latin indices $i, j$ run over $0...3$ for the tangent space of the manifold, while the Greek letters $\mu$, $\nu$ are the coordinate indices on the manifold, also running over $0...3$.\\	
	We consider the action for Myrzakulov $F(R,T)$ gravity \cite{ref56,ref63} as
	\begin{equation}\label{eq3}
		S=\int{e[F(R,T)+L_{m}]}dx^{4},
	\end{equation}
	where $e=\sqrt{-g}$ with $g$ as the determinant of metric tensor $g_{\mu\nu}$, $R=R^{(LC)}+u$ and $T=T^{(W)}+v$ with $R^{(LC)}$ is the Ricci scalar corresponding to Levi-Civita connection and $T^{(W)}$ is the torsion scalar corresponding to Weitzenb\"{o}k connection. And $u$ is a scalar quantity depending on the tetrad, its first and second derivatives, and the connection and its first derivative, and $v$ is is a scalar quantity depending on the tetrad, its first derivative and the connection. Hence, $u$ and $v$ quantify the information on the specific imposed connection \cite{ref63}.\\
	
	The general modified field equations for Myrzakulov $F(R, T)$ gravity are obtained by varying the action \eqref{eq3} with respect to metric field as below (see the reference \cite{ref65} for detail):
	\begin{equation}\label{eq4}
		F_{R}R_{(\mu\nu)}-\frac{1}{2}g_{\mu\nu}F+F_{T}\left(2S_{\nu ij}{S_{\mu}}^{ij}-S_{ij\mu}{S^{ij}}_{\nu}+2S_{\nu ij}{S_{\mu}}^{ji}-4S_{\mu}S_{\nu}\right)=T_{\mu\nu} 
	\end{equation}
	where $F_{R}=\frac{\partial F}{\partial R}, F_{T}=\frac{\partial F}{\partial T}$, $R_{(\mu\nu)}$ is the symmetric part of the Ricci tensor of the affine connection $\Gamma$, ${S_{\mu\nu}}^{\lambda}$ is the torsion tensor, $S_{\mu}$ is the torsion trace, $T_{ij}$ is the stress-energy momentum tensor defined by
	\begin{equation}\label{eq5}
		T_{\mu\nu}=-\frac{2}{\sqrt{-g}}\frac{\delta(\sqrt{-g}L_{m})}{\delta g^{\mu\nu}}
	\end{equation}
	On the other hand, the connection field equations are
	\begin{equation}\label{eq6}
		{P_{\lambda}}^{\mu\nu}(F_{R})+2F_{T}\left({S^{\mu\nu}}_{\lambda}-2{S_{\lambda}}^{[\mu\nu]}-4S^{[\mu}\delta_{\lambda}^{\nu]}\right)=0
	\end{equation}
	where ${P_{\lambda}}^{\mu\nu}(F_{R})$ is the modified Palatini tensor,
	\begin{equation}\label{eq7}
		{P_{\lambda}}^{\mu\nu}(F_{R})=-\frac{\nabla_{\lambda}(\sqrt{-g}F_{R}g^{\mu\nu})}{\sqrt{-g}}+\frac{\nabla_{\alpha}(\sqrt{-g}F_{R}g^{\mu i}\delta_{\lambda}^{\nu})}{\sqrt{-g}}+2F_{R}(S_{\lambda}g^{\mu\nu}-S^{\mu}\delta_{\lambda}^{\nu}-{S_{\lambda}}^{\mu\nu})
	\end{equation}
	with $\nabla$ as the covariant derivative associated with the general affine connection $\Gamma$.\\
	Here, we consider the energy momentum tensor for perfect fluid matter source as
	\begin{equation}\label{eq8}
		T_{\mu\nu}=(\rho+p)U_{\mu}U_{\nu}+pg_{\mu\nu}
	\end{equation}
	where $\rho$, $p$ are respectively, energy density and pressure of the considered perfect fluid matter source, $U^{\mu}=(-1, 0, 0, 0)$ is the four velocity vector.\\
	
	In this paper, we restrict ourselves to the case $u=u(a,\dot{a})$ and $v=v(a,\dot{a})$. The scale factor $a(t)$, the curvature scalar $R$ and the torsion scalar $T$ are taken as independent dynamical variables. Then after some algebra the action \eqref{eq3} becomes \cite{ref72},
	
	\begin{equation}\label{eq9}
		S=\int{Ldt},
	\end{equation}
	where the point-like Lagrangian is given by
	\begin{equation}\label{eq10}
		L=a^{3}(F-TF_{T}-RF_{R}+vF_{T}+uF_{R})-6(F_{R}+F_{T})a\dot{a}^{2}-6(F_{RR}\dot{R}+F_{RT}\dot{T})a^{2}\dot{a}-a^{3}L_{m}.
	\end{equation}
	The corresponding field equations of Myrzakulov $F(R,T)$ gravity are obtained in \cite{ref72,ref73}, as
	\begin{equation}\label{eq11}
		3H\dot{R}F_{RR}-3(\dot{H}+H^{2})F_{R}+3H\dot{T}F_{RT}+6H^{2}F_{T}+\frac{1}{2}F-\frac{1}{2}\dot{a}u_{\dot{a}}F_{R}-\frac{1}{2}\dot{a}v_{\dot{a}}F_{T}=\rho,
	\end{equation}
	
	\begin{multline}\label{eq12}
		\dot{R}^{2}F_{RRR}+(\ddot{R}+2\dot{R}H)F_{RR}+(3H^{2}+2\dot{H}-\frac{1}{2}R)(F_{R}+F_{T})+2\dot{T}HF_{TT}+2\dot{R}\dot{T}F_{RRT}+\dot{T}^{2}F_{RTT}\\
		+(2\dot{R}H+2\dot{T}H+\ddot{T})F_{RT}+\frac{1}{2}F-\frac{1}{6}au_{\dot{a}}\dot{R}F_{RR}-(\frac{1}{2}\dot{a}u_{\dot{a}}+\frac{1}{6}a\dot{u}_{\dot{a}}-\frac{1}{2}u-\frac{1}{6}au_{a})F_{R}-\frac{1}{6}av_{\dot{a}}\dot{T}F_{TT}\\
		-(\frac{1}{2}\dot{a}v_{\dot{a}}+\frac{1}{6}a\dot{v}_{\dot{a}}-\frac{1}{2}v-\frac{1}{6}av_{a})F_{T}-\frac{1}{6}a(v_{\dot{a}}\dot{R}+u_{\dot{a}}\dot{T})F_{RT}=-p.
	\end{multline}
	The energy conservation equation is obtained as
	\begin{equation}\label{eq13}
		\dot{\rho}+3H(\rho+p)=0.
	\end{equation}
	
	%=====================================================================================================================
	\section{Cosmological solutions for $F(R,T)=R+\lambda T$}
	%=====================================================================================================================
	
	In this investigation, we take the arbitrary function $F(R,T)$ in linear form in $R$ and $T$ as given by
	\begin{equation}\label{eq14}
		F(R,T)=R+\lambda T,
	\end{equation}
	where $\lambda$ is an arbitrary constant, $R=u+6(\dot{H}+2H^{2})$ and $T=v-6H^{2}$. Using Eq.~\eqref{eq14} in Eqs.~\eqref{eq11} \& \eqref{eq12}, we obtain the field equations in the form
	\begin{equation}\label{eq15}
		3(1+\lambda)H^{2}+0.5[(u-\dot{a}u_{\dot{a}})+\lambda(v-\dot{a}v_{\dot{a}})]=\rho,
	\end{equation}
	and
	\begin{equation}\label{eq16}
		(1+\lambda)(2\dot{H}+3H^{2})+0.5[u-\dot{a}u_{\dot{a}}-\frac{1}{3}a\dot{u}_{\dot{a}}+\frac{1}{3}au_{a}]+0.5\lambda[v-\dot{a}v_{\dot{a}}-\frac{1}{3}a\dot{v}_{\dot{a}}+\frac{1}{3}av_{a}]=-p.
	\end{equation}
	Now, we consider the scalars $u$ and $v$ in the form of \cite{ref70}
	\begin{equation}\label{eq17}
		u=c_{1}\frac{\dot{a}}{a}\ln\dot{a},~~~~~~v=s(a)\dot{a},
	\end{equation}
	where $c_{1}$ is an arbitrary constant and $s(a)$ is an arbitrary function of scale factor $a$.\\
	Using Eq.~\eqref{eq17} in Eqs.~\eqref{eq15} \& \eqref{eq16}, we get the following form of the field equations \eqref{eq15} \& \eqref{eq16}:
	\begin{equation}\label{eq18}
		3(1+\lambda)H^{2}-\frac{1}{2}c_{1}H=\rho,
	\end{equation}
	\begin{equation}\label{eq19}
		(1+\lambda)(2\dot{H}+3H^{2})-\frac{1}{2}c_{1}H-\frac{1}{6}c_{1}\frac{\dot{H}}{H}=-p.
	\end{equation}
	For $\lambda=0, c_{1}=0$, the field equations, \eqref{eq18} \& \eqref{eq19} will reduced into original Einstein's field equations in general relativity (GR). One can obtain the Friedmann like equations as
	\begin{equation}\label{eq20}
		3H^{2}=\rho+\rho_{MG},
	\end{equation}
	\begin{equation}\label{eq21}
		2\dot{H}+3H^{2}=-p-p_{MG},
	\end{equation}
	where $\rho_{MG}, p_{MG}$ are the geometrical corrections in energy density and pressure, respectively given by
	\begin{equation}\label{eq22}
		\rho_{MG}=\frac{1}{2}c_{1}H-3\lambda H^{2}, ~~~~~~ p_{MG}=-\left[\frac{1}{2}c_{1}H+\frac{1}{6}c_{1}\frac{\dot{H}}{H}-\lambda(2\dot{H}+3H^{2})\right].
	\end{equation}
	These, geometrical corrections, respectively, in energy density and pressure $\rho_{MG}, p_{MG}$, called as effective dark energy sector in Myrzakulov gravity. We can show that effective dark energy sector is conserved, namely $\dot{\rho}_{MG}+3H(\rho_{MG}+p_{MG})=0$, and it can be easily deduced from  matter energy conservation equation $\dot{\rho}+3H(\rho+p)=0$.\\
	We define the matter equation of state as $p=\omega\rho$ with $\omega=$constant and using Eqs.~\eqref{eq13} \& \eqref{eq14}, we get
	\begin{equation}\label{eq23}
		\frac{12(1+\lambda)H-c_{1}}{6H}\dot{H}+\frac{6(1+\lambda)H^{2}-c_{1}H}{2}(1+\omega)=0,
	\end{equation}
	or
	\begin{equation}\label{eq24}
		\frac{12(1+\lambda)H-c_{1}}{6(1+\lambda)H^{2}-c_{1}H}dH+3(1+\omega)\frac{da}{a}=0.
	\end{equation}
	After integration Eq.~\eqref{eq24}, we get
	\begin{equation}\label{eq25}
		6(1+\lambda)H^{2}-c_{1}H-c_{2}\left(\frac{a_{0}}{a}\right)^{3(1+\omega)}=0,
	\end{equation}
	where $c_{2}$ is an integrating constant.\\
	Solving Eq.~\eqref{eq25} for Hubble parameter $H$, we obtain
	\begin{equation}\label{eq26}
		H(a)=\frac{c_{1}}{12(1+\lambda)}+\frac{1}{12}\sqrt{\left(\frac{c_{1}}{1+\lambda}\right)^{2}+\left( \frac{24c_{2}}{1+\lambda}\right)\left(\frac{a_{0}}{a}\right)^{3(1+\omega)}},~~~~~~\lambda\ne-1.
	\end{equation}
	For $c_{1}=0$, we get Hubble parameter as $H(a)=\frac{\sqrt{6c_{2}}}{6\sqrt{1+\lambda}}\left(\frac{a_{0}}{a}\right)^{3(1+\omega)/2}$ which gives a power-law expansion cosmology with a constant deceleration parameter (DP). If we take $c_{2}=0$, then we find $H=$constant which gives exponential-law expansion cosmology with constant DP.\\
	Using the relation $\frac{a_{0}}{a}=1+z$ \cite{ref3}, we get
		\begin{equation}\label{eq27}
		H(z)=\frac{c_{1}}{12(1+\lambda)}+\frac{1}{12}\sqrt{\left(\frac{c_{1}}{1+\lambda}\right)^{2}+\left( \frac{24c_{2}}{1+\lambda}\right)(1+z)^{3(1+\omega)}},~~~~~~\lambda\ne-1.
	\end{equation}
	The deceleration parameter is derived from $q=-1+(1+z)\frac{H'}{H}$ as
	\begin{equation}\label{eq28}
		q(z)=-1+\frac{36(1+\omega)c_{2}(1+z)^{3(1+\omega)}}{\frac{c_{1}^{2}}{1+\lambda}+24c_{2}(1+z)^{3(1+\omega)}+c_{1}\sqrt{\left( \frac{c_{1}}{1+\lambda}\right)^{2}+\frac{24c_{2}}{1+\lambda}(1+z)^{3(1+\omega)}}},~~~~~~\lambda\ne-1.
	\end{equation}
	
	%============================================================================================================
	\section{Observational Constraints}
	%============================================================================================================
	
	For our model and dataset combination, we use the freely available emcee program, available at \cite{ref74}, to conduct an MCMC (Monte Carlo Markov Chain) analysis so that we may compare the model with observational datasets. Through parameter value variation across a variety of cautious priors and analysis of the parameter space posteriors, the MCMC sampler constrains the model and cosmological parameters. We then obtain the one-dimensional and two-dimensional distributions for each parameter: the one-dimensional distribution represents the posterior distribution of the parameter, whilst the two-dimensional distribution shows the covariance between two different values.
	
	%=================================================================
	\subsection{Hubble Function}
	%=================================================================
	
    To ensure the model's validity and feasibility, a model that aligns with observational datasets must be obtained. As a result, in order to obtain this condition, we first investigated $32$ observed statistically non-correlated Hubble datasets $H(z)$ across redshift $z$, with $H(z)$ \cite{ref75}-\cite{ref82} having errors (see Table 1). We used the following $\chi^{2}$-test formula while fitting data: 
	
	\begin{equation}\nonumber
		\chi^{2}(c_{1}, c_{2}, \lambda, \omega)=\sum_{i=1}^{i=N}\frac{[(H_{ob})_{i}-(H_{th})_{i}]^{2}}{\sigma_{i}^{2}}
	\end{equation}
	Where $N$ denotes the total amount of data, $H_{ob},~H_{th}$, respectively, the observed and hypothesized datasets of $H(z)$ and standard deviations are displayed by $\sigma_{i}$.
	%%%%%%%%%%%%%%%%%%%%%%%%%%%%%%%%%%%%%%%%%%%%%%%%%%%%%%%%%%%%%%%%%%%%%%
	%%%%%%%%%%%%%%%%%%%%%%%%%%%%%%%%%%%%%%%%%%%%%%%%%%%%%%%%%%%%%%%%%%%%%%%%%
	\begin{table}[H]
		\centering
 \begin{tabular}{|c|c|c|c|c|}
				\hline
				``S.No. & $z$   & $H(z)$  & $\sigma_{H}$  & Reference\\
				\hline
				1  & $0.07$   & $69.0$    & $19.6$     & \cite{ref75}\\
				2  & $0.09$   & $69.0$    & $12.0$     & \cite{ref76}\\
				3  & $0.12$   & $68.6$    & $26.2$     & \cite{ref75}\\
				4  & $0.17$   & $83.0$    & $8.0$      & \cite{ref76}\\
				5  & $0.179$  & $75.0$    & $4.0$      & \cite{ref77}\\
				6  & $0.199$  & $75.0$    & $5.0$      & \cite{ref77}\\
				7  & $0.2$    & $72.9$    & $29.6$     & \cite{ref75}\\
				8  & $0.27$   & $77.0$    & $14.0$     & \cite{ref76}\\
				9  & $0.28$   & $88.8$     & $36.6$    & \cite{ref75}\\
				10  & $0.352$  & $83.0$    & $14.0$     & \cite{ref77}\\
				11  & $0.3802$ & $83.0$    & $13.5$     & \cite{ref78}\\
				12  & $0.4$    & $95.0$    & $17.0$     & \cite{ref76}\\
				13  & $0.4004$  & $77.0$    & $10.2$   & \cite{ref78}\\
				14  & $0.4247$  & $87.1$    & $11.2$   & \cite{ref78}\\
				15  & $0.4497$  & $92.8$    & $12.9$   & \cite{ref78}\\
				16  & $0.47$    & $89.0$    & $50.0$   & \cite{ref79}\\
				17  & $0.4783$  & $80.9$    & $9.0$    & \cite{ref78}\\
				18  & $0.48$    & $97.0$    & $62.0$   & \cite{ref80}\\
				19  & $0.593$   & $104.0$   & $13.0$   & \cite{ref77}\\
				20  & $0.68$    & $92.0$    & $8.0$    & \cite{ref77}\\
				21  & $0.75$    & $98.8$    & $33.6$   & \cite{ref81}\\
				22  & $0.781$   & $105.0$   & $12.0$   & \cite{ref77}\\
				23  & $0.875$   & $125.0$   & $17.0$   & \cite{ref77}\\
				24  & $0.88$    & $90.0$    & $40.0$   & \cite{ref80}\\
				25  & $0.9$     & $117.0$   & $23.0$   & \cite{ref76}\\
				26  & $1.037$   & $154.0$   & $20.0$   & \cite{ref77}\\
				27  & $1.3$     & $168.0$   & $17.0$   & \cite{ref76}\\
				28  & $1.363$   & $160.0$   & $33.6$   & \cite{ref82}\\
				29  & $1.43$    & $177.0$   & $18.0$   & \cite{ref76}\\
				30  & $1.53$    & $140.0$   & $14.0$   & \cite{ref76}\\
				31  & $1.75$    & $202.0$   & $40.0$   & \cite{ref76}\\
				32  & $1.965$   & $186.0$   & $50.4$   & \cite{ref82}"\\ 
				\hline
		\end{tabular}
		\caption{Observed values of $H(z)$.}\label{T1}
	\end{table}
	%%%%%%%%%%%%%%%%%%%%%%%%%%%%%%%%%%%%%%%%%%%%%%%%%%%%%%%%%%%%%%%%%%%%%%%%%
	
	%%%%%%%%%%%%%%%%%%%%%%%%%%%%%%%%%%%%%%%%%%%%%%%%%%%%%%%%%%%%
	%%%%%%%%%%%%%%%%%%%%%%%%%%%%%%%%%%%%% Figure 1
	%%%%%%%%%%%%%%%%%%%%%%%%%%%%%%%%%%%%%%%%%%%%%%%%%%%%%%%%%%%%
	\begin{figure}[H]
		\centering
		\includegraphics[width=10cm,height=10cm,angle=0]{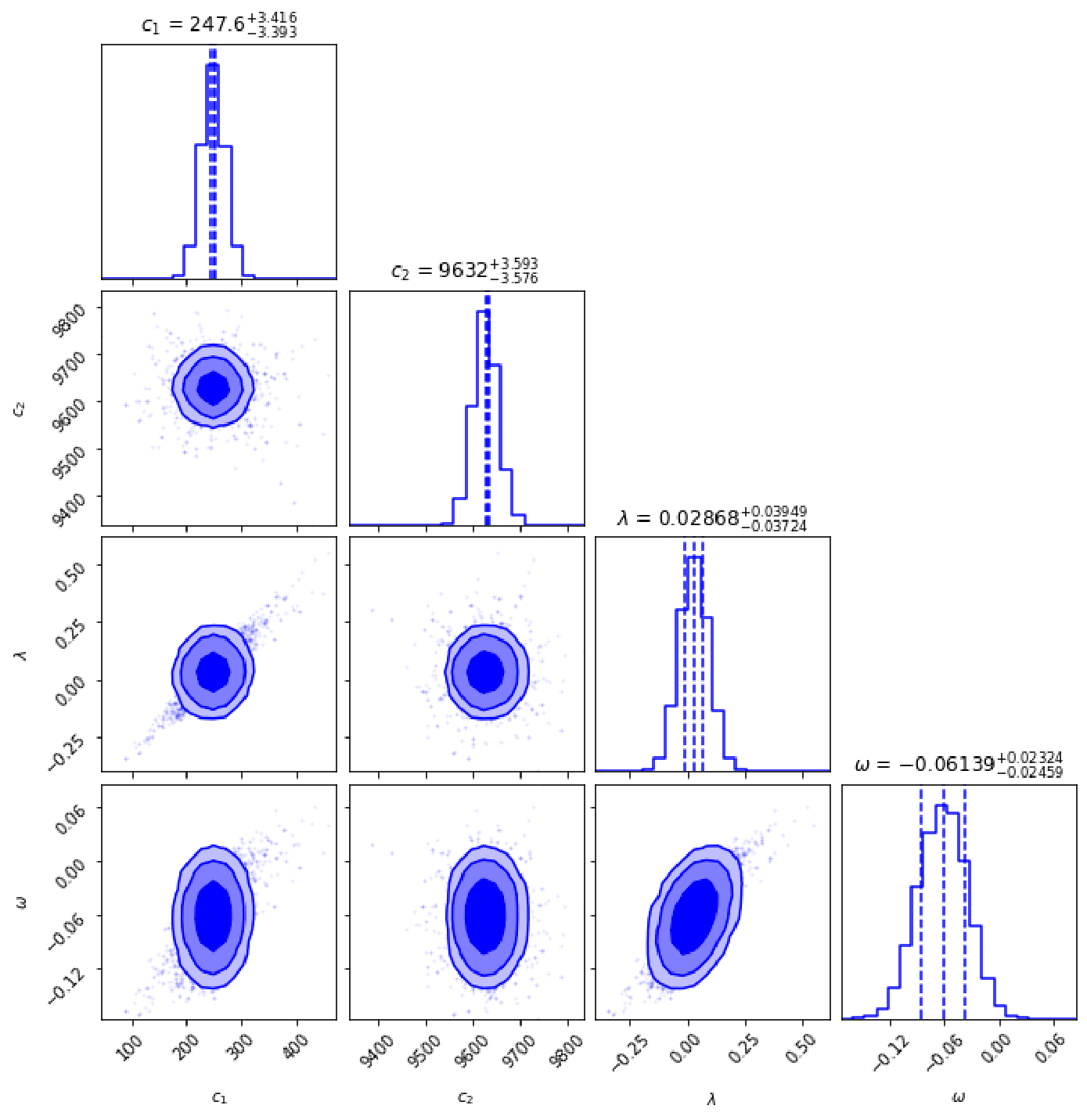}
		\caption{The contour plots of $c_{1}, c_{2}, \lambda, \omega$ at $1-\sigma, 2-\sigma$ and $3-\sigma$ confidence level in MCMC analysis of $H(z)$ datasets.}
	\end{figure}
	%%%%%%%%%%%%%%%%%%%%%%%%%%%%%%%%%%%%%%%%%%%%%%%%%%%%%%%%%%%%%%%%%%
		%%%%%%%%%%%%%%%%%%%%%%%%%%%%%%%%%%%%%%%%%%%%%%%%%%%%%%%%%%%%
	%%%%%%%%%%%%%%%%%%%%%%%%%%%%%%%%%%%%% Figure 2
	%%%%%%%%%%%%%%%%%%%%%%%%%%%%%%%%%%%%%%%%%%%%%%%%%%%%%%%%%%%%
	\begin{figure}[H]
		\centering
		\includegraphics[width=6cm,height=8cm,angle=0]{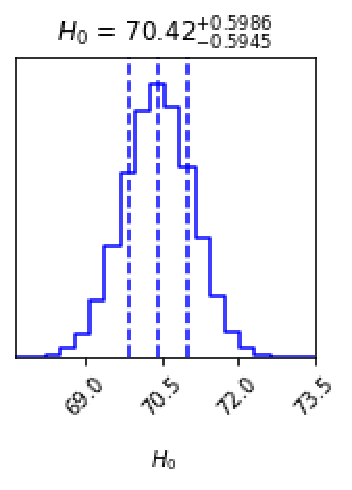}
		\caption{The contour plot of $H_{0}$ at $1-\sigma, 2-\sigma$ and $3-\sigma$ confidence level in MCMC analysis of $H(z)$ datasets for $\Lambda$CDM model.}
	\end{figure}
	%%%%%%%%%%%%%%%%%%%%%%%%%%%%%%%%%%%%%%%%%%%%%%%%%%%%%%%%%%%%%%%%%
	%%%%%%%%%%%%%%%%%%%%%%%%%%%%%%%%%%%%%%%%%%%%%%%%%%%%%%%%%%%%
	%%%%%%%%%%%%%%%%%%%%%%%%%%%%%%%%%%%%% Figure 3
	%%%%%%%%%%%%%%%%%%%%%%%%%%%%%%%%%%%%%%%%%%%%%%%%%%%%%%%%%%%%
	\begin{figure}[H]
		\centering
		a.\includegraphics[width=8cm,height=6cm,angle=0]{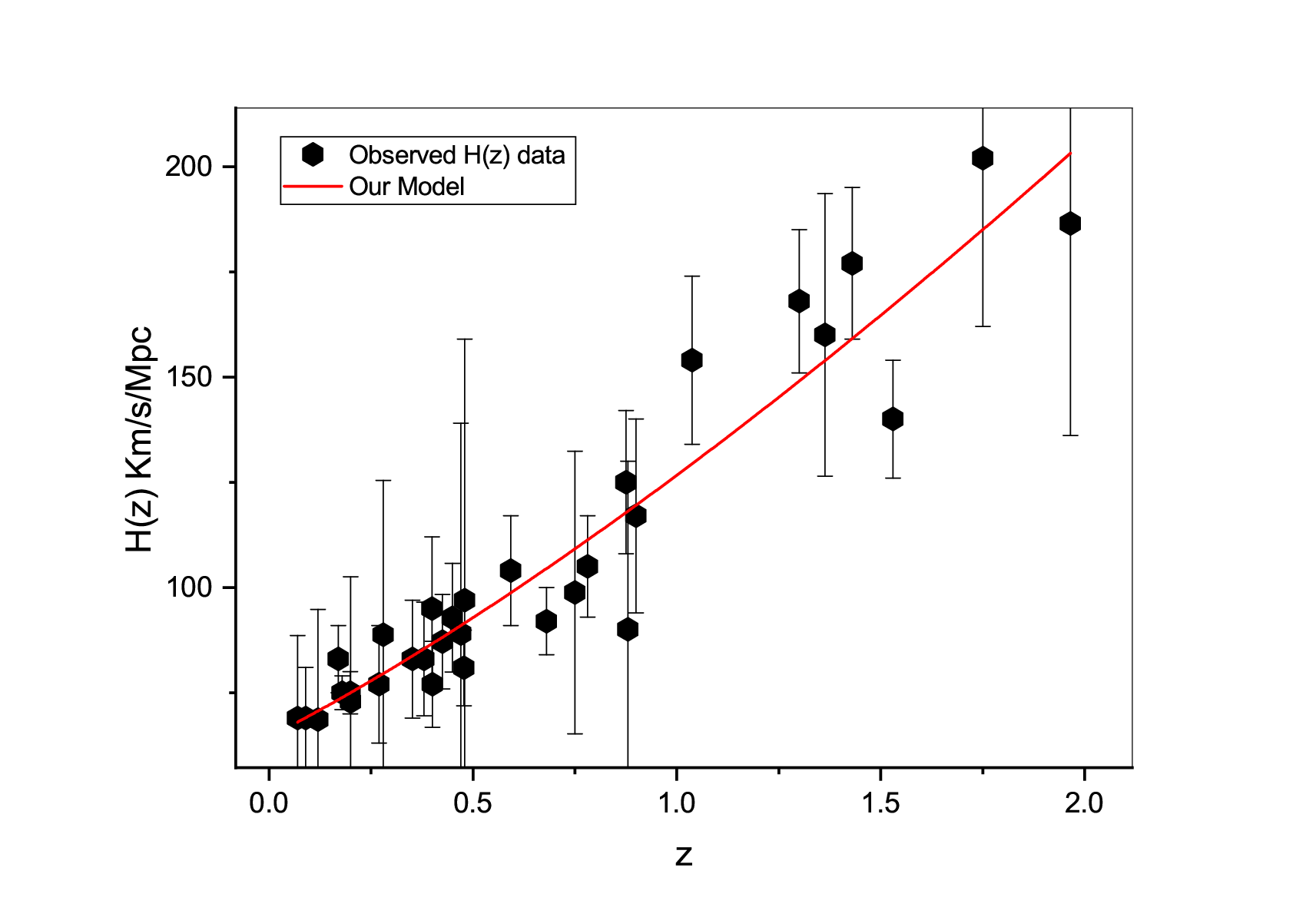}
		b.\includegraphics[width=8cm,height=6cm,angle=0]{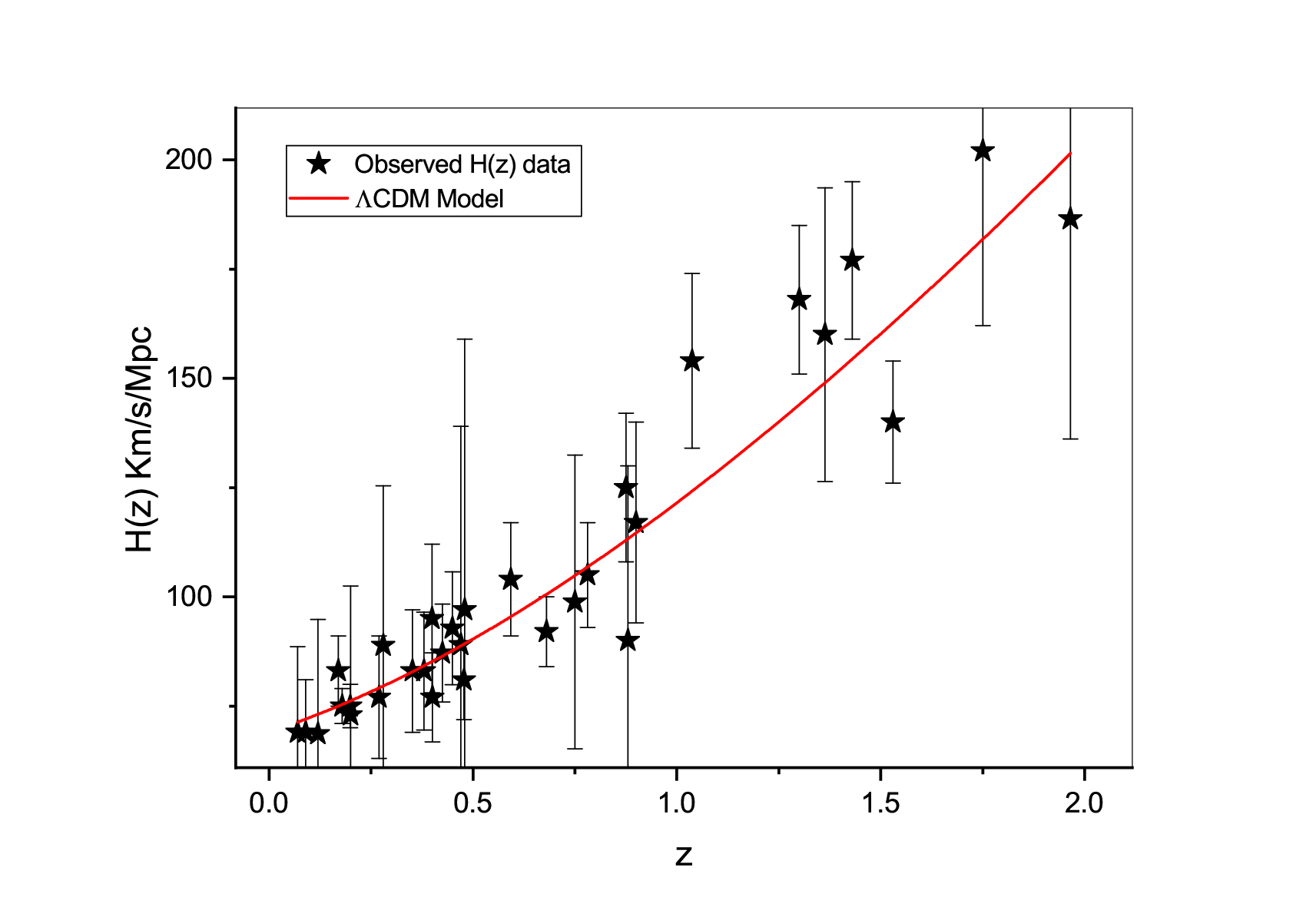}
		\caption{The best fit shape of Hubble parameter $H(z)$ over $z$ for our model and $\Lambda$CDM model with observed non-correlated $H(z)$ datasets, respectively.}
	\end{figure}
	%%%%%%%%%%%%%%%%%%%%%%%%%%%%%%%%%%%%%%%%%%%%%%%%%%%%%%%%%%%%%%%%%%
	%%%%%%%%%%%%%%%%%%%%%%%%%%%%%%%%%%%%%%%%%%%%%%%%%%%%%%%%%%%%%%%%%%%%

	\begin{table}[H]
		\centering
		\begin{tabular}{|c|c|c|c|}
			\hline
						Model                   &Parameter       & Prior            & Value   \\
			\hline
			$\Lambda$CDM Model      & $H_{0}$        & $(50, 100)$      & $70.42_{-0.5945}^{+0.5986}$\\
			\hline
			                        &$c_{1}$         & $(10, 1000)$     & $247.6_{-3.393}^{+3.416}$\\
			                        &$c_{2}$         & $(1000, 10000)$  & $9632_{-3.576}^{+3.593}$\\
			$f(R,T)$-Model& $\lambda$      & $(-1, 1)$        & $0.02868_{-0.03724}^{+0.03949}$\\
			                        &$\omega$        & $(-1, 1)$        & $-0.06139_{-0.02459}^{+0.02324}$\\
			                        &$\chi^{2}$      & --               & $122.69146$\\
			\hline
		\end{tabular}
		\caption{The MCMC Results in $H(z)$ datasets analysis.}\label{T2}
	\end{table}
	%%%%%%%%%%%%%%%%%%%%%%%%%%%%%%%%%%%%%%%%%%%%%%%%%%%%%%%%%%%%%%%%%%%%%%%%%%%%
	For $\Lambda$CDM model, we have considered the Hubble function $H(z)=H_{0}\sqrt{\Omega_{m0}(1+z)^{3}+\Omega_{\Lambda0}}$ with $\Omega_{m0}=0.3$ and $\Omega_{\Lambda0}=0.7$. Using this Hubble function, we have performed the MCMC analysis with $32$ statistically non-correlated Hubble datasets $H(z)$ with error bars in $H(z)$. The output likelihood plot is given in figure 2 and the best fit Hubble curve for $\Lambda$CDM model is shown in figure 3b. We have obtained the best fit value of Hubble constant as $H_{0}=70.42_{-0.5945}^{+0.5986}$ Km/s/Mpc by varying $H_{0}$ in the range $50<H_{0}<100$ for $\Lambda$CDM model which is mentioned in Table 2.\\
	
	The contour plots for parameters $c_{1}, c_{2}, \lambda,$ and $\omega$ at $68\%$, $95\%$, and $99\%$ confidence levels, respectively, are shown in Figure 1. The best fit shape of Hubble function for $F(R, T)$-model with $H(z)$ datasets is shown in figure 3a. As indicated in Table 2, we have selected a broad range of priors for our study in order to estimate the cosmological parameters, which have the highest likelihood of existing for the theoretical values of these parameters for the best-fit model. We have estimated best fit values of $c_{1}=247.6_{-3.393}^{+3.416}$, $c_{2}=9632_{-3.576}^{+3.593}, \lambda=0.02868_{-0.03724}^{+0.03949}$ and $\omega=-0.06139_{-0.02459}^{+0.02324}$ at $1-\sigma, 2-\sigma$ and $3-\sigma$ errors using the priors $(10, 1000)$, $(1000, 10000)$, $(-1, 1)$ and $(-1, 1)$, respectively. We have estimated the Hubble constant as $H_{0}=64.3627_{-1.3408}^{+1.3291}~Km s^{-1} Mpc^{-1}$ for the best fit model. Recently, Cao and Ratra \cite{ref83} have obtained the value of Hubble constant $H_{0}=69.8\pm1.3~Km s^{-1} Mpc^{-1}$ while in \cite{ref84} they estimated this value as $H_{0}=69.7\pm1.2~Km s^{-1} Mpc^{-1}$. Recently, Alberto Dom\'{\i}nguez et al. \cite{ref85} have obtained this parameter in their likelihood analysis of wide observational datasets as $H_{0}=66.6\pm1.6~Km s^{-1} Mpc^{-1}$ and \cite{ref86,ref87} have obtained as $H_{0}=65.8\pm3.4~Km s^{-1} Mpc^{-1}$. Freedman et al. \cite{ref88} have estimated the present value of Hubble constant $H_{0}=69.6\pm0.8~Km s^{-1} Mpc^{-1}$, Birrer et al. \cite{ref89} have measured $H_{0}=67.4_{-3.2}^{+4.1}~Km s^{-1} Mpc^{-1}$, Boruah et al. \cite{ref90} have measured $H_{0}=69_{-2.8}^{+2.9}~Km s^{-1} Mpc^{-1}$ and most recently, Freedman \cite{ref91} has estimated $H_{0}=69.8\pm0.6~Km s^{-1} Mpc^{-1}$ and Qin Wu et al. \cite{ref92} have measured $H_{0}=68.81_{-4.33}^{+4.99}~Km s^{-1} Mpc^{-1}$. Recently, in 2018 \cite{ref93}, the Plank Collaboration estimated that the Hubble constant is currently $H_{0}=67.4\pm0.5$ km/s/Mpc, while Riess et al. \cite{ref94} obtained $H_{0}=73.2\pm1.3$ km/s/Mpc in 2021. In comparison of the above results, the result obtained in our model for $H_{0}$ is compatible with observational datasets.
	
	%============================================================
	\subsection{Apparent Magnitude $m(z)$}
	%============================================================
	
    The relationship between luminosity distance and redshift is one of the main observational techniques used to track the universe's evolution. The expansion of the cosmos and the redshift of the light from distant brilliant objects are taken into consideration when calculating the luminosity distance ($D_{L}$) in terms of the cosmic redshift ($z$). It is provided as
	\begin{equation}\label{eq29}
		D_{L}=a_{0} r (1+z),
	\end{equation}
	where the radial coordinate of the source $r$, is established by
	\begin{equation}\label{eq30}
		r  =  \int^r_{0}dr = \int^t_{0}\frac{cdt}{a(t)} = \frac{1}{a_{0}}\int^z_0\frac{cdz}{H(z)},
	\end{equation}
	where we have used $ dt=dz/\dot{z}, \dot{z}=-H(1+z)$.\\
	As a result, the luminosity distance is calculated as follows:
	\begin{equation}\label{eq31}
		D_{L}=c(1+z)\int^z_0\frac{dz}{H(z)}.
	\end{equation}
	Hence, the apparent magnitude $m(z)$ of a supernova is defined as:
	\begin{equation}\label{eq32}
		m(z)=16.08+ 5~log_{10}\left[\frac{(1+z)H_{0}}{0.026} \int^z_0\frac{dz}{H(z)}\right].
	\end{equation}
	We use the most recent collection of $1048$ datasets of the Pantheon SNe Ia samples in the ($0.01 \le z \le 1.7$) range \cite{ref95} in our MCMC analysis. We have used the following $\chi^{2}$ formula to constrain different model parameters:
	\begin{equation}\nonumber
		\chi^{2}(c_{1}, c_{2}, \lambda, \omega, H_{0})=\sum_{i=1}^{i=N}\frac{[(m_{ob})_{i}-(m_{th})_{i}]^{2}}{\sigma_{i}^{2}}.
	\end{equation}
	The entire amount of data is denoted by $N$, the observed and theoretical datasets of $m(z)$ are represented by $m_{ob}$ and $m_{th}$, respectively, and standard deviations are denoted by $\sigma_{i}$.\\
	
	%%%%%%%%%%%%%%%%%%%%%%%%%%%%%%%%%%%%%%%%%%%%%%%%%%%%%%%%%%%%%%%%%%%%%%%%%%%%%%%%%%%%%%%%
	\begin{table}[H]
		\centering
		\begin{tabular}{|c|c|c|}
			\hline
			
			Parameter             & Prior          & Value\\
			\hline
			$c_{1}$         & $(10, 1000)$     & $294_{-0.1110}^{+0.1125}$\\
			$c_{2}$         & $(1000, 10000)$  & $9625_{-0.1131}^{+0.1121}$\\
			$\lambda$       & $(-1, 1)$        & $0.01452_{-0.01326}^{+0.02639}$\\
			$\omega$        & $(-1, 1)$        & $0.02946_{-0.05842}^{+0.04105}$\\
			$H_{0}$         & $(50, 100)$      & $67.85_{-0.1272}^{+0.1104}$\\
			$\chi^{2}$            &  --            & $5430.38409$  \\
			\hline
		\end{tabular}
		\caption{The MCMC Results in Pantheon SNe Ia datasets analysis.}\label{T3}
	\end{table}
	%%%%%%%%%%%%%%%%%%%%%%%%%%%%%%%%%%%%%%%%%%%%%%
	%%%%%%%%%%%%%%%%%%%%%%%%%%%%%%%%%%%%%%%%%%%%%%%%%%%%%%%%%%%%%%%%%%%
	
	The mathematical expression for apparent magnitude $m(z)$ is represented in Eq.~\eqref{eq32} and figure 4 shows the contour plots for $c_{1}, c_{2}, \lambda, \omega, H_{0}$ in MCMC analysis of Pantheon SNe Ia datasets. Figure 5 depict the best fit curve of apparent magnitude versus $z$ for Pantheon SNe Ia datasets for the best fit values of model parameters. We have applied a wide range priors $(10, 1000), (1000, 10000), (-1, 1), (-1, 1), (50, 100)$ for $c_{1}, c_{2}, \lambda, \omega, H_{0}$, respectively, in our analysis and obtained the best fit values as $c_{1}=294_{-0.1110}^{+0.1125}, c_{2}=9625_{-0.1131}^{+0.1121}, \lambda=0.01452_{-0.01326}^{+0.02639}, \omega=0.02946_{-0.05842}^{+0.04105}, H_{0}=67.85_{-0.1272}^{+0.1104}$ with  $1-\sigma$, $2-\sigma$ \& $3-\sigma$ errors at $68\%$, $95\%$ \& $99\%$ confidence level, respectively (see Table 3). Our result is compatible with the recent observational datasets.
	%%%%%%%%%%%%%%%%%%%%%%%%%%%%%%%%%%%%%%%%%%%%%%%%%%%%%%%%%%%%
	%%%%%%%%%%%%%%%%%%%%%%%%%%%%%%%%%%%%% Figure 4
	%%%%%%%%%%%%%%%%%%%%%%%%%%%%%%%%%%%%%%%%%%%%%%%%%%%%%%%%%%%%
	\begin{figure}[H]
		\centering
		\includegraphics[width=10cm,height=10cm,angle=0]{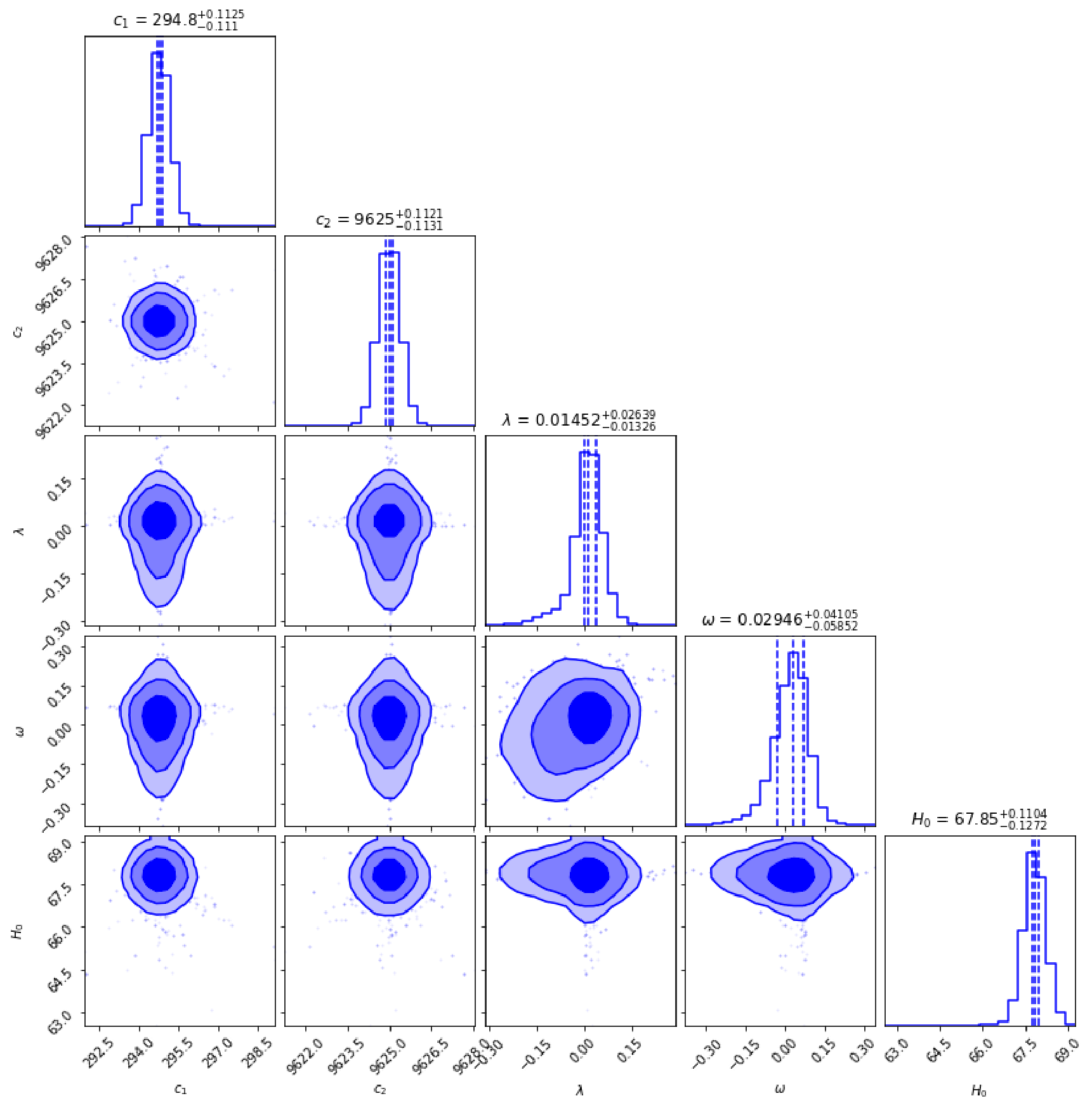}
		\caption{The contour plots of $c_{1}, c_{2}, \lambda, \omega, H_{0}$ in MCMC analysis of the Pantheon SNe Ia samples.}
	\end{figure}
	%%%%%%%%%%%%%%%%%%%%%%%%%%%%%%%%%%%%%%%%%%%%%%%%%%%%%%%%%%%%%%%%%%%
		%%%%%%%%%%%%%%%%%%%%%%%%%%%%%%%%%%%%%%%%%%%%%%%%%%%%%%%%%%%%
	%%%%%%%%%%%%%%%%%%%%%%%%%%%%%%%%%%%%% Figure 5
	%%%%%%%%%%%%%%%%%%%%%%%%%%%%%%%%%%%%%%%%%%%%%%%%%%%%%%%%%%%%
	\begin{figure}[H]
		\centering
		\includegraphics[width=10cm,height=7cm,angle=0]{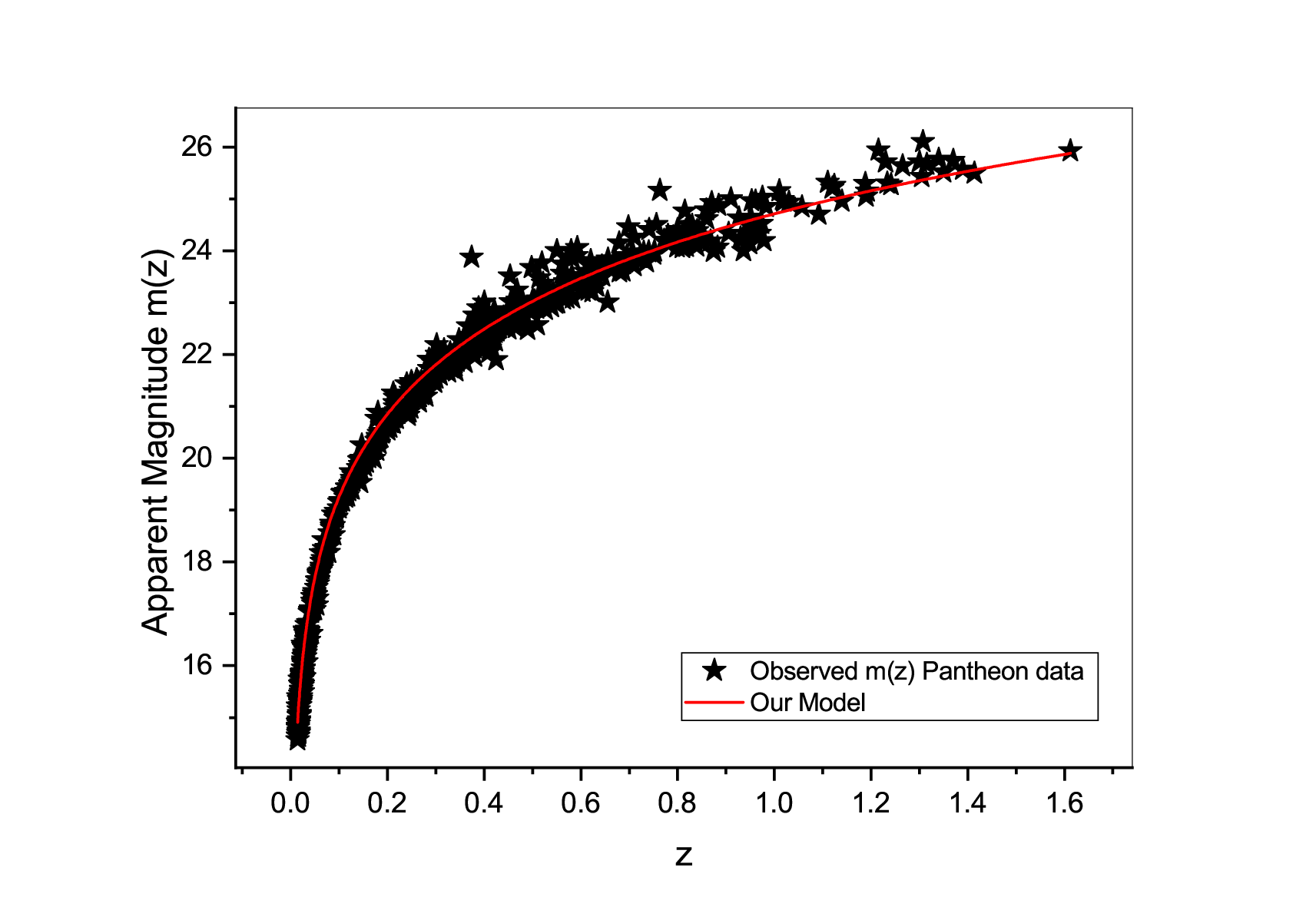}
		\caption{The best fit plot of apparent magnitude $m(z)$ versus $z$ for Pantheon SNe Ia samples.}
	\end{figure}
	%%%%%%%%%%%%%%%%%%%%%%%%%%%%%%%%%%%%%%%%%%%%%%%%%%%%%%%%%%%%%%%%%%%
	
	%========================================================================================
    \section{Result Analysis and Discussion}
    %========================================================================================
    
    In this section, first we introduce matter energy density parameter $\Omega_{m}$ and effective dark energy density parameter $\Omega_{MG}$, respectively as
        \begin{equation}\label{eq33}
    	\Omega_{m}=\frac{\rho}{3(1+\lambda)H^{2}}, ~~~~~~ \Omega_{MG}=\frac{c_{1}}{6(1+\lambda)H}.
    \end{equation}
    From Eq.~\eqref{eq18}, we can define the relationship between energy density parameters $\Omega_{m}$ \& $\Omega_{MG}$ as
    \begin{equation}\label{eq34}
    	\Omega_{m}+\Omega_{MG}=1.
    \end{equation}
    Equation \eqref{eq33} represents the expressions for matter energy density parameter $\Omega_{m}$ and effective dark energy density parameter $\Omega_{MG}$, respectively. The geometrical evolution of $\Omega_{m}$, $\Omega_{MG}$, respectively are shown in figure 6a \& 6b. Figure 6a depicts that the early universe is matter dominated $\lim_{z\to\infty}\Omega_{m}\to1$ and in late-time universe $\lim_{z\to-1}\Omega_{m}\to0$. Figure 6b depicts that late-time universe is dark energy dominated $\lim_{z\to-1}\Omega_{MG}\to1$ and in early time universe $\lim_{z\to\infty}\Omega_{MG}\to0$. At present $z=0$, we have estimated values of these parameters as $(\Omega_{m0}, \Omega_{MG0})=(0.3767_{-0.0559}^{+0.0620}, 0.6233_{-0.0018}^{+0.0016}), (0.2831_{-0.0156}^{+0.0294}, 0.7169_{-0.0041}^{+0.0020})$, respectively, along two observational datasets $H(z)$ and Pantheon SNe Ia datasets while for standard $\Lambda$CDM model, these quantities are as $\Omega_{m0}=0.30$ and $\Omega_{\Lambda0}=0.70$ with Hubble constant $H_{0}=70.42_{-0.5945}^{+0.5986}$ Km/s/Mpc. These values are compatible with recent observations  \cite{ref83}-\cite{ref88}. The good observations in our model are the effective dark energy term that comes from the geometrical corrections.
    
    %%%%%%%%%%%%%%%%%%%%%%%%%%%%%%%%%%%%%%%%%%%%%%%%%%%%%%%%%%%%
    %%%%%%%%%%%%%%%%%%%%%%%%%%%%%%%%%%%%% Figure 6
    %%%%%%%%%%%%%%%%%%%%%%%%%%%%%%%%%%%%%%%%%%%%%%%%%%%%%%%%%%%%
    \begin{figure}[H]
    	\centering
    	a.\includegraphics[width=8cm,height=7cm,angle=0]{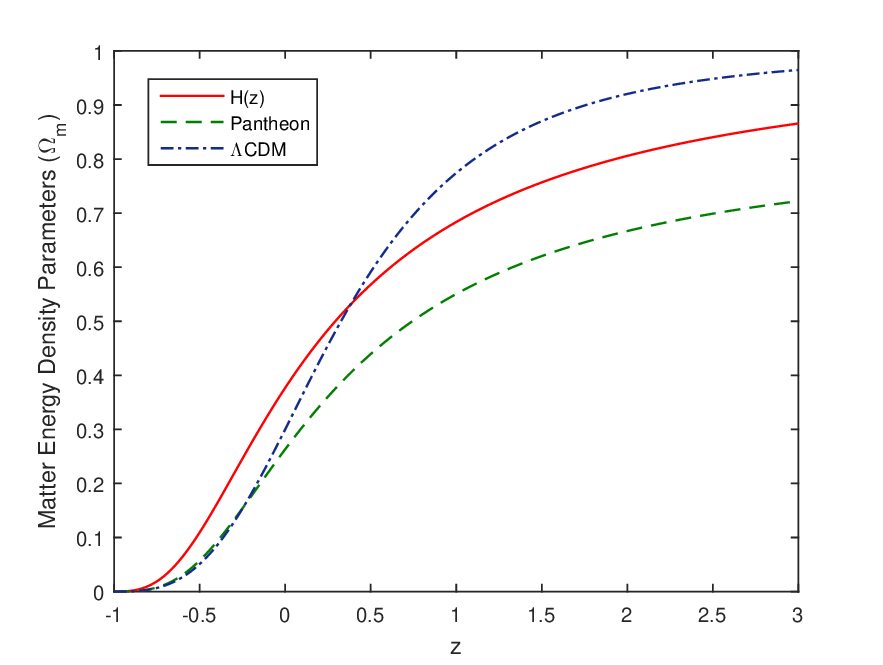}
    	b.\includegraphics[width=8cm,height=7cm,angle=0]{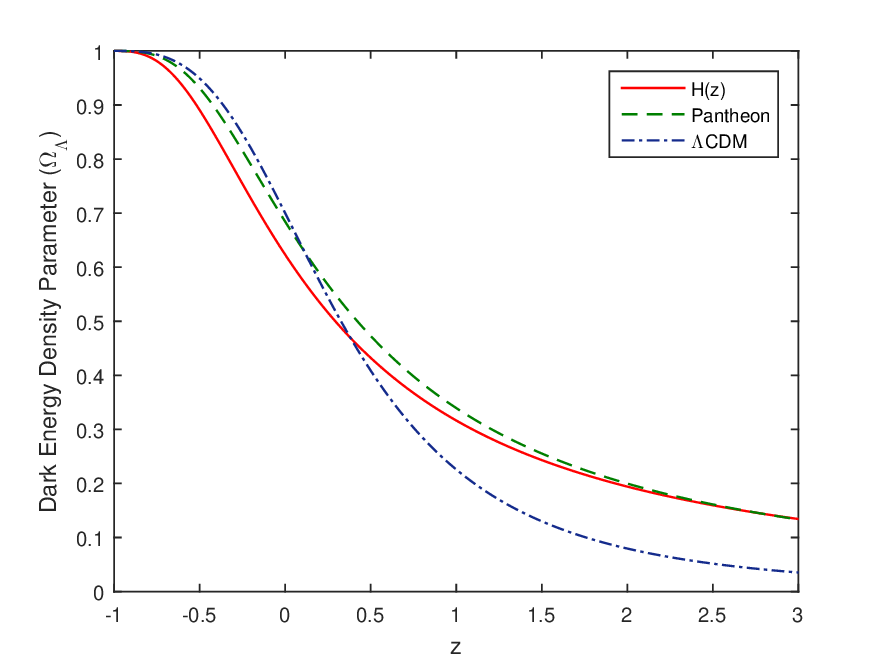}
    	\caption{The geometrical evolution of matter energy density parameter $\Omega_{m}$ and effective dark energy density parameter $\Omega_{MG}=\Omega_{\Lambda}$ over $z$, respectively.}
    \end{figure}
    %%%%%%%%%%%%%%%%%%%%%%%%%%%%%%%%%%%%%%%%%%%%%%%%%%%%%%%%%%%%%%%%%%
    
    The effective dark energy equation of state parameter $\omega_{de}$ is obtained as
    \begin{equation}\label{eq35}
    	\omega_{de}=-1+\frac{1+z}{3}\frac{H'}{H}-\frac{2\lambda(1+z)H'}{c_{1}-6\lambda H},
    \end{equation}
    or
    	\begin{multline}\label{eq36}
    	\omega_{de}(z)=-1+\frac{12(1+\omega)c_{2}(1+z)^{3(1+\omega)}}{\frac{c_{1}^{2}}{1+\lambda}+24c_{2}(1+z)^{3(1+\omega)}+c_{1}\sqrt{\left( \frac{c_{1}}{1+\lambda}\right)^{2}+\frac{24c_{2}}{1+\lambda}(1+z)^{3(1+\omega)}}}\\
    	-\frac{6\lambda c_{2}(1+\omega)(1+z)^{3(1+\omega)}}{\frac{c_{1}(2+\lambda)}{2}\sqrt{\left( \frac{c_{1}}{1+\lambda}\right)^{2}+\frac{24c_{2}}{1+\lambda}(1+z)^{3(1+\omega)}}-\frac{\lambda(1+\lambda)}{2}\left[\left( \frac{c_{1}}{1+\lambda}\right)^{2}+\frac{24c_{2}}{1+\lambda}(1+z)^{3(1+\omega)}\right] },~~~~~~\lambda\ne-1.
    \end{multline}
    
    %%%%%%%%%%%%%%%%%%%%%%%%%%%%%%%%%%%%%%%%%%%%%%%%%%%%%%%%%%%%
    %%%%%%%%%%%%%%%%%%%%%%%%%%%%%%%%%%%% Figure 7
    %%%%%%%%%%%%%%%%%%%%%%%%%%%%%%%%%%%%%%%%%%%%%%%%%%%%%%%%%%%%
    \begin{figure}[H]
    	\centering
    	\includegraphics[width=10cm,height=8cm,angle=0]{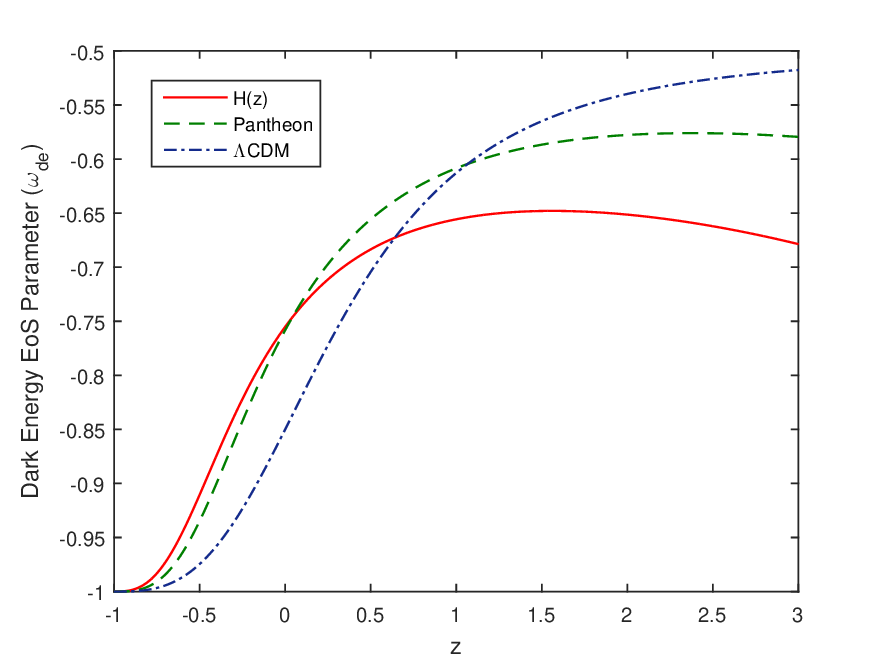}
    	\caption{The evolution of effective dark energy equation of state parameter $\omega_{de}$ versus $z$.}
    \end{figure}
    %%%%%%%%%%%%%%%%%%%%%%%%%%%%%%%%%%%%%%%%%%%%%%%%%%%%%%%%%%%%%%%%%%
    
    The mathematical expression for effective dark energy EoS parameter $\omega_{de}$ is represented in Eq.~\eqref{eq36} and its geometrical behaviour is shown in figure 7. From figure 7, we can see that effective dark energy EoS varies as $-1\le\omega_{de}\le-0.6787$ along $H(z)$ datasets, $-1\le\omega_{de}\le-0.5795$ along Pantheon datasets and $-1\le\omega_{de}\le-0.5176$ for $\Lambda$CDM model, over the redshift $-1\le z \le 3$. At $z=0$, we have measured the value of EoS $\omega_{de}=-0.7552_{-0.0109}^{+0.0079}, 0.7583_{-0.0018}^{+0.0103}$, respectively, along two observational datasets and for $\Lambda$CDM model it is estimated as $\omega_{de}=-0.85$. Also from figure 7, we observe that $\omega_{de}\to-1$ as $z\to-1$ (at late-time universe) for all datasets. Thus, these behaviours of effective dark energy EoS parameter $\omega_{de}$ confirm that our model is in good agreement with observational datasets, and our derived $F(R,T)$ model is very closed to $\Lambda$CDM standard cosmological model.\\
    
    The expression for deceleration parameter $q(z)$ is represented in Eq.~\eqref{eq28} and its geometrical nature is depicted in figure 8. From figure 8, we observe that $\lim_{z\to-1}q\to-1$ (accelerating phase of late-time universe) and $\lim_{z\to\infty}q\to\frac{1+3\omega}{2}>0$ (decelerating phase of early universe) that reveals that for to obtain past decelerating universe the perfect fluid equation of state parameter should be $\omega>-\frac{1}{3}$. At present ($z=0$) we have estimated the value of deceleration parameter $q_{0}=-0.2295_{-0.0226}^{+0.0218}, -0.2590_{-0.0455}^{+0.0372}$, respectively along two observational datasets $H(z)$ and Pantheon SIe Ia and for $\Lambda$CDM standard model, it is obtained as $q_{0}=-0.55$, and this reveals that the present phase of the expanding universe is accelerating which is in good agreement with recent observations. From figure 8, one can see that evolution of $q(z)$ shows a signature-flipping (transition) point called as transition redshift $z_{t}$ at which $q=0$ i.e., the expansion of universe is in accelerating phase for $z<z_{t}$ and it is in decelerating expansion phase for $z>z_{t}$. The general expression for $z_{t}$, we have derived from Eq.~\eqref{eq28} as below
    \begin{equation}\label{eq37}
    	z_{t}=\left[ \frac{c_{1}^{2}(2+3\omega)}{6c_{2}(1+\lambda)(1+3\omega)}\right]^{\frac{1}{3(1+\omega)}}-1,~~~~~~\lambda\ne-1.
    \end{equation}
    In the derived model, we have measured the transition redshift as $z_{t}=0.4438_{-0.790}^{+0.1008}, 0.3651_{-0.0904}^{+0.1644}$, respectively, for two observational datasets $H(z)$ and Pantheon SNe Ia, while for standard $\Lambda$CDM model, it is obtained as $z_{t}=0.671$. From Eq.~\eqref{eq32}, we can obtain ever accelerating universe for $\omega\to-1$ as $z_{t}\to\infty$ Recently in 2013, Farooq and Ratra \cite{ref96} have measured this decelerating-accelerating transition redshifts $z_{t}=0.74\pm0.05$ while Farooq et al. \cite{ref97} have estimated as $z_{t}=0.74\pm0.04$. In 2016, Farooq et al. \cite{ref98} have measured this transition redshifts $z_{t}=0.72\pm0.05$ and in 2018, Yu et al. \cite{ref99} have suggested this transition redshifts varies over $0.33 < z_{t} < 1.0$. Thus, the decelerating-accelerating transition redshift $z_{t}=0.4438_{-0.790}^{+0.1008}, 0.3651_{-0.0904}^{+0.1644}$ measured in our model is in good agreement with the results obtained in \cite{ref96}-\cite{ref101}.
    
%%%%%%%%%%%%%%%%%%%%%%%%%%%%%%%%%%%%%%%%%%%%%%%%%%%%%%%%%%%%
%%%%%%%%%%%%%%%%%%%%%%%%%%%%%%%%%%%% Figure 8
%%%%%%%%%%%%%%%%%%%%%%%%%%%%%%%%%%%%%%%%%%%%%%%%%%%%%%%%%%%%
\begin{figure}[H]
	\centering
	\includegraphics[width=10cm,height=8cm,angle=0]{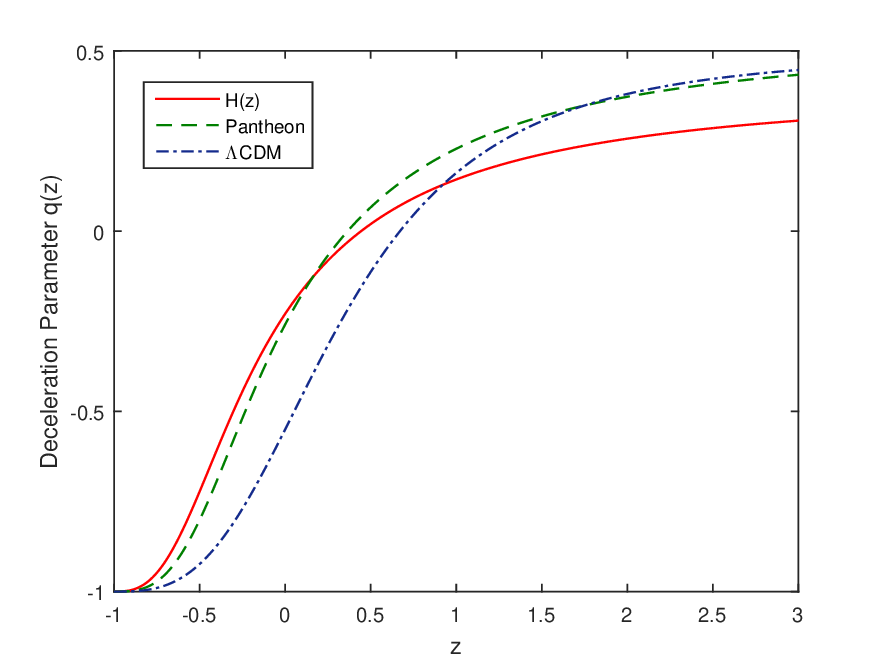}
	\caption{The geometrical evolution of deceleration parameter $q(z)$ versus $z$.}
\end{figure}
%%%%%%%%%%%%%%%%%%%%%%%%%%%%%%%%%%%%%%%%%%%%%%%%%%%%%%%%%%%%%%%%%%

    %===================================================================================
	\subsection*{Om diagnostic analysis}
	%===================================================================================
	
	It is simpler to classify concepts related to cosmic dark energy because of the behavior of Om diagnostic function \cite{ref102}. For a spatially homogeneous universe, the Om diagnostic function is given as
	\begin{equation}\label{eq38}
		Om(z)=\frac{\left(\frac{H(z)}{H_{0}}\right)^{2}-1}{(1+z)^{3}-1},~~~~z\ne0
	\end{equation}
	where $H_{0}$ denotes the current value of the Hubble parameter $H(z)$ as stated in Eq.~\eqref{eq27}. A negative slope of $Om(z)$ indicates quintessence motion, whereas a positive slope denotes phantom motion. The $\Lambda$CDM model is represented by the constant $Om(z)$.\\	
	Using Eq.~\eqref{eq27} in \eqref{eq38}, we get
	\begin{equation}\label{eq39}
		Om(z)=  \frac{\left(\left[\frac{c_{1}}{12(1+\lambda)}+\frac{1}{12}\sqrt{\left(\frac{c_{1}}{1+\lambda}\right)^{2}+\left( \frac{24c_{2}}{1+\lambda}\right)(1+z)^{3(1+\omega)}}\right]/H_{0}\right)^{2}-1}{(1+z)^{3}-1},~~~~z\ne0
	\end{equation}
	The mathematical expression for $Om(z)$ function is represented in Eq.~\eqref{eq39} and its geometrical behaviour is shown in figure 9. From figure 9, we observe that the slopes are negative along the both datasets $H(z)$ datasets and Pantheon SNe Ia datasets, during evolution of the universe and hence, our model behaves just like quintessence dark energy model. At late-time $\lim_{z\to-1}Om(z)\to\left[1-\frac{c_{1}^{2}}{36(1+\lambda)^{2}H_{0}^{2}}\right]$ which is a constant and it indicates that our model tends to $\Lambda$CDM model in late-time scenario.
	
	%%%%%%%%%%%%%%%%%%%%%%%%%%%%%%%%%%%%%%%%%%%%%%%%%%%%%%%%%%%%
	%%%%%%%%%%%%%%%%%%%%%%%%%%%%%%%%%%%% Figure 9
	%%%%%%%%%%%%%%%%%%%%%%%%%%%%%%%%%%%%%%%%%%%%%%%%%%%%%%%%%%%%
	\begin{figure}[H]
		\centering
		\includegraphics[width=10cm,height=8cm,angle=0]{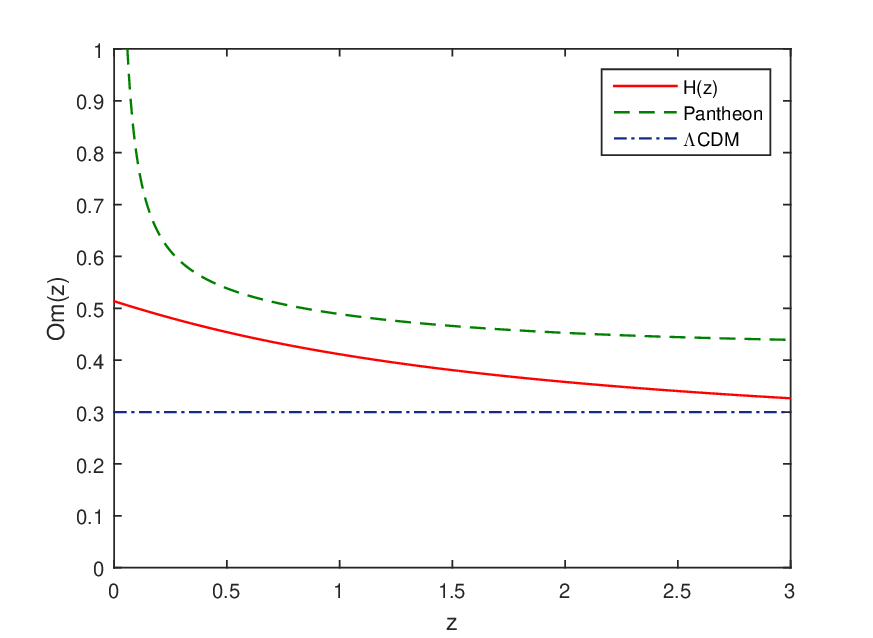}
		\caption{Evolution of $Om(z)$ parameter versus $z$.}
	\end{figure}
	%%%%%%%%%%%%%%%%%%%%%%%%%%%%%%%%%%%%%%%%%%%%%%%%%%%%%%%%%%%%%%%%%%
	
	%=============================================================================================
	\section{Age of the Universe}
	%=============================================================================================
	
	We define the age of the universe as
	\begin{equation}\label{eq40}
		t_{0}-t=\int_{0}^{z}\frac{dz}{(1+z)H(z)},~~~~z\ge0
	\end{equation}
	where $H(z)$ is given by Eq.~\eqref{eq27}. Using this in \eqref{eq40}, we have
	\begin{equation}\label{eq41}
	   (t_{0}-t)=\lim_{z\to\infty}\int_{0}^{z}\frac{dz}{(1+z)\left[\frac{c_{1}}{12(1+\lambda)}+\frac{1}{12}\sqrt{\left(\frac{c_{1}}{1+\lambda}\right)^{2}+\left( \frac{24c_{2}}{1+\lambda}\right)(1+z)^{3(1+\omega)}}\right]}.
	\end{equation}
		We can see that as $z\to\infty$, $(t_{0}-t)$ tends to a constant value that represents the cosmic age of the universe, $(t_{0}-t)\to t_{0}=0.0141602_{-0.0000655}^{+0.0001027}, 0.0122838_{-0.0002805}^{+0.0006345}$, respectively, along two datasets $H(z)$ and Pantheon SNe Ia. The present cosmic age of the universe, we have measured as $t_{0}=13.8486_{-0.0640}^{+0.1005}, 12.0135_{-0.2743}^{+0.6206}$ Gyrs, respectively along two observational datasets, which are very closed to observational estimated values and estimated $\Lambda$CDM value $t_{0}=13.3895_{-0.1129}^{+0.1240}$ Gyrs. Recently \cite{ref103,ref104} have measured present age of the universe as $t_{0}\approx13.87$ Gyrs.
	
	%=============================================================================================
	\section{Conclusions}
	%=============================================================================================
	
	    We study exact cosmological models in Myrzakulov $F(R,T)$ gravity theory in the current paper. The arbitrary function $F(R, T)=R+\lambda T$ has been investigated, in which $R$ represents the Ricci-scalar curvature, $T$ is the torsion scalar, and $\lambda$ is an arbitrary constant. After solving the field equations in a flat FLRW spacetime manifold for the Hubble parameter, we estimated the best fit values of the model parameters with $1-\sigma, 2-\sigma$, and $3-\sigma$ regions by utilizing the MCMC analysis. We have conducted a model discussion and outcome analysis using these best fit model parameter conditions. For the best fit shape of Hubble function $H(z)$, we have found the values of model parameters as $c_{1}=247.6_{-3.393}^{+3.416}$, $c_{2}=9632_{-3.576}^{+3.593}, \lambda=0.02868_{-0.03724}^{+0.03949}$ and $\omega=-0.06139_{-0.02459}^{+0.02324}$ at $1-\sigma, 2-\sigma$ and $3-\sigma$ errors for $H(z)$ datasets, and $c_{1}=294.8_{-0.1110}^{+0.1125}, c_{2}=9625_{-0.1131}^{+0.1121}, \lambda=0.01452_{-0.01326}^{+0.02639}, \omega=-0.02946_{-0.05852}^{+0.04105}, H_{0}=67.85_{-0.1272}^{+0.1104}$ with  $1-\sigma$, $2-\sigma$ \& $3-\sigma$ errors at $68\%$, $95\%$ \& $99\%$ confidence level, respectively, for Pantheon SNe Ia datasets (see Table 2 \& 3). We have also find the best fit value of Hubble constant for $\Lambda$CDM model with statistically non-correlated $H(z)$ datasets as $H_{0}=70.42_{-0.5945}^{+0.5986}$ Km/s/Mpc. In the analysis of deceleration parameter $q(z)$, our universe model shows a transit phase dark energy model that is decelerating $q>0$ for $z>z_{t}$ and accelerating $q<0$ for $z<z_{t}$. We have found the transition redshift $z_{t}=0.4438_{-0.790}^{+0.1008}, 0.3651_{-0.0904}^{+0.1644}$, respectively for two observational datasets $H(z)$ and Pantheon. We have found the present value of DP as $q_{0}=-0.2295_{-0.0226}^{+0.0218}, -0.2590_{-0.0455}^{+0.0372}$ with Hubble constant extcolor{red}{$H_{0}=64.3627_{-1.3408}^{+1.3291}, 67.85_{-0.1272}^{+0.1104}~Km s^{-1} Mpc^{-1}$, respectively, for two datasets. The Om diagnostic analysis of $H(z)$ indicates that the current behaviour of our model is quintessential and late-time it approaches to $\Lambda$CDM model. We have found that $(\Omega_{m}, \Omega_{MG})\to(0, 1)$ at late-time which is good observations for our model. We have found the present values of total energy density parameters as $(\Omega_{m0}, \Omega_{MG0})=(0.3767_{-0.0559}^{+0.0620}, 0.6233_{-0.0018}^{+0.0016}), (0.2831_{-0.0156}^{+0.0294}, 0.7169_{-0.0041}^{+0.0020})$, respectively, along two observational datasets $H(z)$ and Pantheon SNe Ia datasets while for standard $\Lambda$CDM model, these quantities are as $\Omega_{m0}=0.30$ and $\Omega_{\Lambda0}=0.70$. From figure 7, we have found that the effective dark energy EoS parameter varies as $-1\le\omega_{de}\le-0.6787$ along $H(z)$ datasets, $-1\le\omega_{de}\le-0.5795$ along Pantheon datasets and $-1\le\omega_{de}\le-0.5176$ for $\Lambda$CDM model, over the redshift $-1\le z \le 3$. At $z=0$, we have measured the value of EoS $\omega_{de}=-0.7552_{-0.0109}^{+0.0079}, 0.7583_{-0.0018}^{+0.0103}$, respectively, along two observational datasets and for $\Lambda$CDM model it is estimated as $\omega_{de}=-0.85$ with $\omega_{de}\to-1$ as $z\to-1$ at late-time universe. We have found the present age of the universe for our derived $F(R,T)$ model as $t_{0}=13.8486_{-0.0640}^{+0.1005}, 12.0135_{-0.2743}^{+0.6206}$ Gyrs, respectively along two observational datasets, which are very closed to observational estimated values and estimated $\Lambda$CDM value
	    $t_{0}=13.3895_{-0.1129}^{+0.1240}$ Gyrs.\\
	    	    
	    Thus, we have found that the above derived $F(R,T)$ gravity model can describe the accelerated phase of expanding universe without introducing the dark energy term $\Lambda$. The results of $F(R,T)$ gravity model is extremely very similar and closed to $\Lambda$CDM standard cosmological model but without introducing cosmological constant $\Lambda$-term. Also, we can recover the original Friedmann model without $\Lambda$-term from $F(R,T)$ gravity model by substituting $\lambda=0, c_{1}=0$. This $F(R,T)$ gravity theory is the generalization of both $F(R)$ and $F(T)$ gravity theory. Thus, the present modified gravity model is interesting and attracts to researcher in this field to re-investigate it for exploring the hidden cosmological properties of this $F(R,T)$ gravity theory.                                                                                                                                             
	    \section{Acknowledgments}
			
			This work was supported by the Ministry of Science and Higher Education of the Republic of Kazakhstan, Grant AP14870191. The authors are thankful to renowned Reviewers/Editors for their valuable suggestion to improve this manuscript.
%%%%%%%%%%%%%%%%%%%%%%%%%%%%%%%%%%%%%%%%%%%%%%%%%%%%%%%%%%%%%%%%%%%%%%%%%%%%%%%%%%%%%%%%%%%%%%%%%%%%
%%%%%%%%%%%%%%%%%%%%%%%%%%%%%%%%%%%%%%%%%%%%%%%%%%%%%%%%%%%
%%%%%%%%%%%%%%%%%%%%%%%%%%%%%%%%%%%%%%%%%%%%%%%%%%%%%%%%%%%%%%%%%%%%%%%%%%%%%%%%%%%%%%%%%%%%%%%%%%%%%%
\section{Statements and Declarations}
\subsection*{Funding and/or Conflicts of interests/Competing interests}
The author of this article has no conflict of interests. The author have no competing interests to declare that are relevant to the content of this article. Authors have mentioned clearly all received support from the organization for the submitted work.
    %%%%%%%%%%%%%%%%%%%%%%%%%%%%%%%%%%%%%%%%%%%%%%%%%%%
		
\end{document}